\documentclass[5p,times]{elsarticle}

\journal{Chemistry and Physics of Lipids}

\usepackage{graphicx}
\usepackage{amsmath}
\usepackage{url}

\begin{document}

\hyphenation{phos-pho-lip-id phos-pho-cho-line}

\begin{frontmatter}

\title{Growth and shape transformations of giant phospholipid
  vesicles upon interaction with an aqueous oleic acid suspension}

\author[ibf]{Primo\v{z} Peterlin\corref{pp}}
\cortext[pp]{Corresponding author. Fax: +386-1-4315127.} 
\ead{primoz.peterlin@mf.uni-lj.si}
\author[ibf]{Vesna Arrigler}
\author[fkkt]{Ksenija Kogej}
\author[ibf,ijs]{Sa\v{s}a Svetina}
\author[eth]{Peter Walde}

\address[ibf]{University of Ljubljana, Faculty of Medicine, Institute of
  Biophysics, Lipi\v{c}eva 2, SI-1000 Ljubljana, Slovenia}
\address[fkkt]{University of Ljubljana, Faculty of Chemistry and
  Chemical Technology, A\v{s}ker\v{c}eva 5, SI-1000 Ljubljana,
  Slovenia}
\address[ijs]{Jo\v{z}ef Stefan Institute, Jamova 39, SI-1000 Ljubljana,
  Slovenia}
\address[eth]{ETH Z\"{u}rich, Department of Materials,
  Wolfgang-Pauli-Str. 10, CH-8093 Z\"{u}rich, Switzerland}

\begin{abstract}
  The interaction of two types of vesicle systems was investigated:
  micrometer-sized, giant unilamellar vesicles (GUVs) formed from
  1-palmitoyl-2-oleoyl-\emph{sn}-glycero-3-phosphocholine (POPC) and
  submicrometer-sized, large unilamellar vesicles (LUVs) formed from
  oleic acid and oleate, both in a buffered aqueous solution
  ($\mathrm{pH}=8.8$).  Individual POPC GUVs were transferred with a
  micropipette into a suspension of oleic acid/oleate LUVs, and the
  shape changes of the GUVs were monitored using optical microscopy.
  The behavior of POPC GUVs upon transfer into a 0.8~mM suspension of
  oleic acid, in which oleic acid/oleate forms vesicular bilayer
  structures, was qualitatively different from the behavior upon
  transfer into a 0.3~mM suspension of oleic acid/oleate, in which
  oleic acid/oleate is predominantly present in the form of monomers
  and possibly non-vesicular aggregates.  In both cases,
  changes in vesicle morphology were observed within tens of seconds
  after the transfer.  After an initial increase of the vesicle
  cross-section, the vesicle started to evaginate, spawning dozens of
  satellite vesicles connected to the mother vesicle with narrow necks
  or tethers.  In 60\% of the cases of transfer into a 0.8~mM oleic
  acid suspension, the evagination process
  reversed and proceeded to the point where the membrane formed
  invaginations.  In some of these cases, several consecutive
  transitions between invaginated and evaginated shapes were observed.
  In the remaining 40\% of the cases of transfer into the 0.8~mM oleic
  acid suspension and in all cases of vesicle transfer into the 0.3~mM
  oleic acid suspension, no invaginations nor subsequent evaginations
  were observed.  An interpretation of the observed vesicle shape
  transformation on the basis of the bilayer-couple model is
  proposed, which takes into account uptake of oleic
    acid/oleate molecules by the POPC vesicles, oleic acid flip-flop
    processes and transient pore formation.
\end{abstract}

\end{frontmatter}


\section{Introduction}

The paper presents a study of the interaction between a
micrometer-sized giant unilamellar vesicle (GUV) made from
1-palmitoyl-2-oleoyl-\emph{sn}-glycero-3-phosphocholine (POPC) and a
buffered (pH~8.8) suspension of oleic acid/oleate, consisting of
either 100~nm large unilamellar vesicles (LUVs) at a concentration
of oleic acid exceeding the critical concentration for vesicle
formation (cvc), or predominantly monomers at concentrations below the
cvc.  The interaction was studied through the growth and the
morphological shape changes of the POPC GUV, as visualized by an
optical microscope.

Certain single-chain amphiphiles, \emph{e.g.}, fatty acids with 10
carbon atoms or more, are known to form bilayers when their
concentration is above the cvc, the temperature of the suspension is
above the Krafft temperature (\emph{i.e.,} the temperature at which
their solubility equals the cvc), and its pH is above 7
\cite{Gebicki:1976,Hargreaves:1978}.  Their self-assembly differs from
that of double-chain amphiphiles in a number of ways. On one hand, the
cvc for single-chain amphiphiles is much higher ($\sim
10^{-3}$--$10^{-4}$~M) than for phospholipids (\emph{e.g.}, the
value for 1,2-dipalmitoyl-\emph{sn}-glycero-3-phosphocholine (DPPC) is
$4.6\times 10^{-10}$~M) \cite{Smith:1972}.  On the other hand,
unlike bilayers formed from phospholipids, which remain stable through
a wide range of pH values, the type of aggregate formed by fatty acids
in aqueous medium depends strongly on the pH of the medium, with two
immiscible phases when the pH is clearly below the (apparent) acid
dissociation constant ($\textrm{pH} \ll \textrm{pK}_a$), bilayers when
$\textrm{pH} \sim \textrm{pK}_a$ and micelles when $\textrm{pH} \gg
\textrm{pK}_a$.

A number of studies exists where similar interactions between
  phospholipid and oleic acid/oleate aggregates were studied on LUVs.
It has been shown previously \cite{Cheng:2003} that upon mixing POPC
vesicles and oleic acid/oleate vesicles buffered to $\textrm{pH} \sim
\textrm{pK}_a$, the oleic acid/oleate vesicles disappear, which has
been explained by the dissociation of oleate from the oleic
acid/oleate vesicles and its uptake by the phosphatidylcholine
vesicles.  In another experiment \cite{Fujikawa:2005}, a
  similar migration of fatty acid molecules from fatty acid vesicles
  to mixed phospholipid/fatty acid micelles has been observed.  In
suspensions of fatty acid vesicles at $\textrm{pH} \sim
\textrm{pK}_a$, Luisi and co-workers as well as others
\cite{Blochliger:1998,Hanczyc:2004,Chungcharoenwattana:2005b} reported
that the presence of existing oleic acid or phospholipid
\cite{Lonchin:1999} vesicles catalyzed the spontaneous formation of
oleic acid vesicles resulting in newly formed vesicles that were more
closely related in size to the pre-formed vesicles than to those
formed spontaneously.  Experiments with ferritin-labeled vesicles
\cite{Berclaz:2001a} indicated that the new vesicles formed via the
process of increase of membrane area and fission of the preformed
vesicles, rather than by being formed \emph{de novo}.  These
experiments were conducted with large unilamellar vesicles, typically
between 50 and 200~nm in size, and the observation relied on dynamic
light scattering and electron microscopy
\cite{Berclaz:2001a,Berclaz:2001b,Stano:2006}.  The former method
allows for monitoring the presence and/or size of vesicles/micelles in
dependence on time, but does not provide a direct visual control,
while the latter method offers a direct visual observation without the
monitoring of the time dependence.

The work presented here tried to overcome the described limitations by
studying the effects of fatty acid incorporation into phospholipid
GUVs in a manner which allowed for a real-time visual monitoring using
optical microscopy.  In the experiments described, a single POPC GUV
was transferred with a micropipette into a chamber filled with a
suspension of oleic acid/oleate vesicles.  For a comparison, a
transfer into a suspension of oleic acid at a concentration lower than
the cvc has also been performed.

The vesicle shape is a very sensitive indicator of the amount of oleic
acid/oleate and/or phospholipid molecules in either leaflet of the
bilayer, since the difference of the areas of the two membrane
leaflets depends on them \cite{Svetina:1989}.  The occupancy of oleic
acid/oleate molecules in either leaf\-let in turn depends on the
balance between the rates of association, dissociation and
translocation (flip-flop) of oleic acid molecules in the phospholipid
membrane.  Similar methodology -- studying the effects of membrane
modification through shape changes of phospholipid GUVs -- has been
used, for instance, in observations of chemically-induced shape
transformations of GUVs \cite{Menger:1992}, in experiments with
lysolipids which intercalate into the phospholipid membrane
\cite{Needham:1995,Tanaka:2004,Inaoka:2007}, in the case when
  flippase activity was induced \cite{Papadopulos:2007}, in
experiments where phospholipase A2 was microinjected onto or into
individual vesicles, thus converting phospholipids into lysolipids
\cite{Wick:1996}, and for monitoring the effects of peptide
incorporation into the phospholipid membrane
\cite{Tamba:2005,Mally:2007}.  An important new feature of this study in relation to
the ones mentioned above is that in the system studied here, during
the course of the reaction, there is a much more significant increase
of the membrane area due to the oleic acid/oleate incorporation into
the membrane. Another important difference is that, unlike
  simple surfactants like lysolipids, which exist in either monomeric
  or micellar state, fatty acids can form bilayer structures.

\section{Materials and methods}

\subsection{Materials}

Oleic acid (\textit{cis}-9-octadecenoic acid), D-(+)-glucose and
D-(+)-sucrose were from Fluka (Buchs, Switzerland); Trizma base and
Trizma HCl were from Sigma-Aldrich (St. Louis, MO, USA).
1-palmitoyl-2-oleoyl-\textit{sn}-glycero-3-phospho\-choline (POPC) was
purchased from Avanti Polar Lipids (Alabaster, AL, USA; purity
$>99$\%, used without further purification).
2-(12-(7-nitrobenz-2-oxa-1,3-diazol-4-yl)amino)dodecanoyl-1-hexa\-decanoyl-\textit{sn}-glycero-3-phospho\-choline
(NBD C${}_{12}$-HPC) was purchased from Invitrogen (Eugene, OR, USA).
Methanol and chloroform were purchased from Kemika (Zagreb, Croatia).
All the solutions were prepared in doubly distilled and sterile water.

\subsection{Preparation of giant lipid vesicles}

Giant lipid vesicles were prepared from POPC using electroformation (a
modified method of Angelova \cite{Angelova:1986}) in 0.2~M sucrose and
diluted with 0.2~M glucose, both buffered with 5~mM Trizma (pH~8.8).
The lipids were dissolved in a mixture of chloroform/methanol (2:1,
v/v) to a concentration of 1~mg/mL.  25~$\mu$L of the lipid solution
was spread onto two Pt electrodes and vacuum-dried for 2 hours.  The
electrodes were then placed into an electroformation chamber
\cite{Angelova:1988}, which was filled with 0.2~M sucrose, buffered
with 5~mM Trizma to pH~8.8. AC current (4~V/10~Hz) was then applied.
After 2~hours the voltage and the frequency were reduced in steps,
first to 3~V/5~Hz, after 15 minutes to 2~V/2.5 Hz, and after
additional 15 minutes to the final values of 1~V/1~Hz, which were held
for 30 minutes.  The chamber was then drained and flushed with
buffered 0.2~M glucose solution, yielding a suspension of GUVs in a
1:1 sucrose/glucose solution. The size distribution of the
  vesicles in the suspension was determined using dynamic light
  scattering on a 3D-DLS Research Lab (LS Instruments, Fribourg,
  Switzerland).  Two samples were taken from a vial with a suspension
  of GUVs: one from the upper fraction, and another one from the
  bottom of the vial after the vesicles were left to settle for
  30~minutes.  Photon correlation spectroscopy showed that size
  distributions of both samples have a peak at 150~nm, the difference
  being that the bottom fraction contains a greater proportion of
  larger vesicles.  As verified by optical microscopy, GUVs containing
  entrapped sucrose were mostly unilamellar and spherical, with
  diameters up to 100~$\mu$m.

GUVs labeled with a fluorescent marker were prepared in a similar
manner.  First, NBD-labeled lipid was dissolved in
chloroform/methanol (2:1, v/v) to a concentration of 1~mg/mL.  Then,
the solution of the NBD-labeled lipid in the chloroform/methanol
mixture was added to a solution of lipid in the chloroform/methanol
mixture in order to obtain either 1 or 2~wt\% NBD-labeled lipid
solution.  The rest of the procedure for GUV preparation was the same
as for the unmarked lipids.

\subsection{Oleic acid suspensions}

A suspension of large unilamellar vesicles (LUVs) from oleic
acid/oleate was prepared using extrusion \cite{MacDonald:1991}. Oleic
acid was dispersed in 0.2~M glucose solution with 5~mM Trizma
buffer (pH~8.8) to a concentration of 50~mM. The suspension was
stirred overnight with a magnetic stirrer bar before being diluted
with 0.2~M glucose prepared with 5~mM Trizma
buffer (pH~8.8) to the preferred concentration (0.3~mM or
0.8~mM). The rationale behind the choice of the oleic acid concentrations
used is that the lower one is assumed to be below the cvc, and the
higher one above it, as the cvc for oleic acid/oleate in the pH range
8.5--9 is estimated to be around 0.4--0.7~mM \cite{Walde:1994b},
even though values as low as 0.082~mM \cite{Chen:2004a} or as high
as 1.0~mM \cite{Hildebrand:2004} have been reported. The diluted
suspension was subjected to 11 passes through two 100~nm polycarbonate
membranes mounted in an Avestin LiposoFast extruder (Ottawa, Canada).

The osmolality of the suspension was determined by a Knauer automatic
cryoscopic semi-micro osmometer (Berlin, Germany). The osmolality of
0.2~M buffered glucose solution was found to be 216~mOsm/kg and
the osmolality of the suspension of 50~mM oleic acid in 0.2~M
buffered glucose solution was found to be 207~mOsm/kg. As the
suspension was diluted with buffered glucose solution before the
experiment in a ratio 1:100 (v/v) or more, the difference between the
osmolality of suspension and the osmolality of the buffered glucose
solution was below the experimental error.


The size of oleic acid/oleate vesicles was determined using dynamic
light scattering on a Malvern Zetasizer 3000 (Malvern Instruments,
Malvern, UK).  A 50~mM oleic acid/oleate suspension was prepared
as described above.  In order to achieve sufficient optical scattering
signal, the suspension was extruded without dilution.  A disposable
cuvette was filled with approximately 2~ml of this suspension.  Photon
correlation spectroscopy showed that the distribution of oleic
acid/oleate vesicle sizes (diameters) at pH~8.8 had a single peak at
$69.1\pm 2.5$~nm, with a polydispersity index \cite{Frisken:2001} of
$0.42\pm 0.14$.

\subsection{Suspension of POPC LUVs}

POPC LUVs were prepared using the freeze-thaw extrusion method.  A
round-bottomed flask was filled with 120.8~$\mu$L of 10~mg/mL POPC
dissolved in a chloroform/methanol (2:1, v/v) mixture, to which
200~$\mu$L of chloroform was added.  By turning the flask, chloroform
evaporated and a lipid film was evenly distributed on the wall of the
flask. Residual organic solvent was removed by drying at a reduced
pressure (60~mmHg, water aspirator) for 2~h.  The lipid film was then
hydrated with 2.1~mL of 0.2~M glucose solution and vortexed using
250~$\mu$m glass beads.  After six freeze-thaw cycles, 2~ml of the
suspension was passed 11 times through two 100~nm polycarbonate
membranes mounted in an Avestin LiposoFast extruder
\cite{MacDonald:1991}.

\begin{figure*}
  \begin{center}
    \begin{tabular}{cc}
      \includegraphics[width=0.48\textwidth]{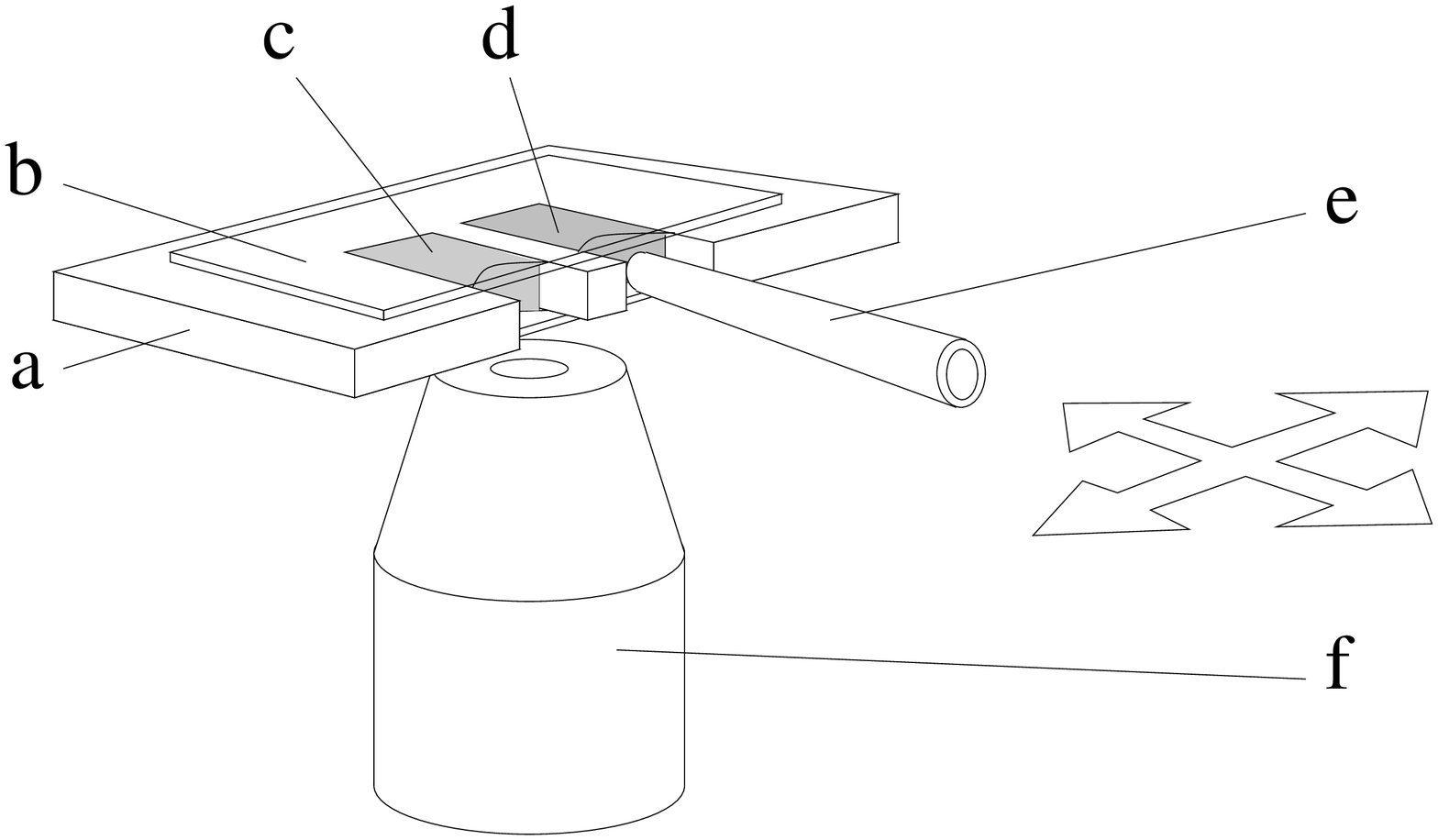} &
      \includegraphics[width=0.20\textwidth]{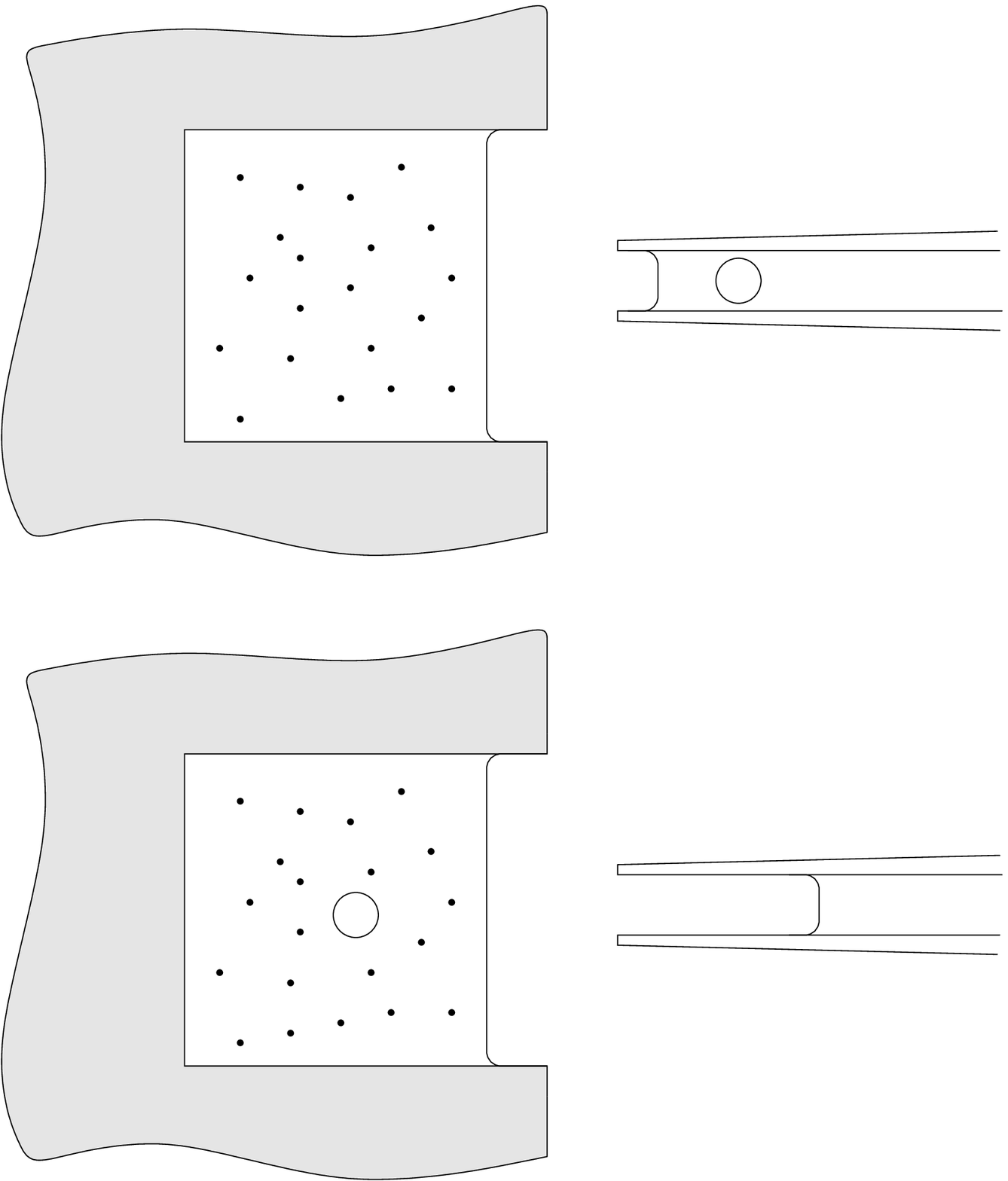} \\
      (A) & (B)
    \end{tabular}
  \end{center}
  \caption{Micromanipulation setup for transferring POPC GUVs into a
    chamber with an oleic acid/oleate suspension (A).  a -- frame
    machined from 2~mm acrylic glass (polymethyl methacrylate, PMMA),
    mounted to the microscope stage; b -- top $32 \times 24$~mm cover
    glass (an identical one is on the bottom side), mounted with a
    vacuum grease; c -- chamber with the suspension of oleic
    acid/oleate LUVs; d -- chamber with the suspension of POPC GUVs; e
    -- micropipette mounted to the micromanipulator; f -- microscope
    objective.  B: Chamber c before (top) and immediately after
    (bottom) transfer of a POPC GUV.  Schemes are not drawn to scale.}
  \label{fig:micromanipulation}
\end{figure*}

\subsection{Microscopy and micromanipulation}

The influence of aqueous suspensions of oleic acid on POPC GUVs was
observed under an inverted optical microscope (Zeiss/Opton IM 35,
objective Plan40/0.60 Ph2; Oberkochen, Germany) at 0.3~mM or
0.8~mM concentrations of oleic acid/oleate in the suspension.  The
micromanipulation chamber used had two compartments; one was filled
with a suspension of POPC GUVs buffered to pH~8.8, the other
(volume $\approx 250$~$\mu$L) with an
oleic acid/oleate suspension at a given oleic acid/oleate
concentration, also buffered to pH~8.8.  A spherical POPC GUV was
selected, fully aspirated into a glass micropipette with a diameter
exceeding the diameter of the vesicle, and transferred into the target
suspension containing oleic acid, where the content of the
micropipette was released (Fig.~\ref{fig:micromanipulation}).  After
removing the micropipette, the vesicle was monitored and its shape
changes were observed until the vesicle shape did not change any more.
The whole process was recorded with a Sony SSC-M370CE B/W video
camera (Sony, Tokyo, Japan) and a Panasonic AG-7350 S-VHS video
recorder (Matsushita Electric Industrial Co., Osaka, Japan).  To allow
for access with a micropipette, both compartments in a
micromanipulation chamber were relatively thick (approximately 3~mm)
and open on one side.  An unwanted consequence of this was a constant
convection flow in the chamber, both because of the evaporation and
because of the heating due to the illumination light.

Fluorescent micrographs were obtained with an inverted optical
microscope (Nikon Diaphot 200, objective Fluor 60/0.70 Ph3DM; Tokyo,
Japan) with the epi-fluorescence attachment, micromanipulating
equipment and a cooled digital camera (Hamamatsu ORCA-ER;
C4742-95-12ERG; Hamamatsu, Japan), connected to a PC running Hamamatsu
Wasabi software. The software also controlled a Uniblitz shutter
(Vincent Associates, Rochester, NY, USA) in the light path of the
Hg-arc light source. Micromanipulation chamber and the transfer
procedure were identical in both setups.

Total internal reflection fluorescence microscopy was performed on an
inverted optical microscope (Nikon Eclipse TE2000-E, objective Plan
Apo TIRF 60/1.45 Oil) with an Ar-laser (488 nm; Melles Griot,
Carlsbad, CA, USA) and a cooled digital camera (Nikon DS-2MBWc),
connected to a PC running LIM Lucia G software (Laboratory Imaging,
Prague, Czech Republic).

In the preparation procedure used, POPC vesicles in the
glucose/sucrose solution mixture containing entrapped sucrose solution
were transferred into an iso-osmolar glucose solution containing oleic
acids.  As the refractive index of a 0.2~M glucose solution
($n=1.3316$, 26${}^\circ$C) differs from the refractive index of a
0.2~M sucrose solution ($n=1.3355$, 26${}^\circ$C), the vesicle
interior appears darker under a phase contrast microscope, and also
exhibits a characteristic white halo around the vesicle.  Refractive
indexes of glucose and sucrose solutions were measured using an Abbe
refractometer (Xintian WY1A, Guiyang, China).

\subsection{Titration of a sodium oleate solution}

A 20~mL sample of 10~mM oleic acid suspension in water was
prepared and an equivalent amount of 1~M sodium hydroxide was
added. 15~mL of this solution was transferred into a thermostated
measuring cell.  The oleate solution was titrated at 25${}^\circ$C
with 0.1~M HCl in a nitrogen atmosphere under continuous stirring.
The pH reached a constant value within ten minutes after the addition
of each portion of the titrant.  The pH was measured with an Iskra MA
5740 pH meter (Ljubljana, Slovenia).

\subsection{Fluorescence spectrometry}

Pyrene fluorescence emission spectra were recorded on a Perkin-Elmer
LS-50 luminescence spectrometer (Norwalk, CT, USA) in a 10~mm quartz
cuvette.  The samples in concentration range from 0.1 to 1.5~mM of
oleic acid were buffered to pH~8.7.  Using excitation at 330~nm, the
emission spectra of pyrene were recorded in the range from 350 to
550~nm.  From the spectra recorded at low oleic acid concentration,
monomer peaks $I_1$ and $I_3$ were identified at 373 and 384~nm,
respectively, while at higher concentrations of oleic acid, an
additional broad excited-state dimer (excimer) peak $I_E$ appeared at
473~nm, indicating formation of vesicles and/or micelles.

\section{Results}
\label{sec:results}

\subsection{General observations and classification of vesicle shape
  transformations}
\label{subsec:classification}

Under an inverted optical microscope, a single GUV prepared from POPC
was transferred with a micropipette from a compartment with a POPC GUV
suspension to a compartment with an oleic acid/oleate LUV suspension.
A total of 54 vesicle transfers into a suspension containing
0.8~mM oleic acid and 9 transfers into a suspension containing
0.3~mM oleic acid were carried out and recorded on video.

\begin{figure}
  \centering\includegraphics[width=0.4\textwidth]{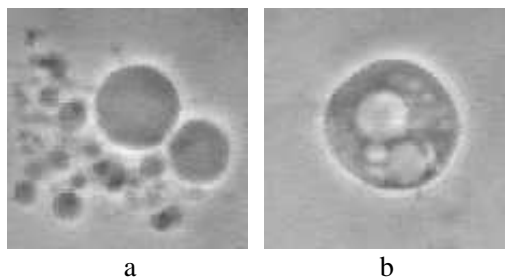}
  \caption{Classification of vesicle transfers with respect to the
    vesicle morphological changes in time course.  Each group is
    labeled based on the sequence of evaginations (E) and
    invaginations (I) exhibited by the vesicle after the transfer into
    an oleic acid suspension, \emph{e.g.}, EI denotes a group of
    vesicle transfers which first exhibit vesicle evagination (a),
    being followed by an invagination of the vesicle (b).  A sequence
    of morphological shapes with typical vesicle shape for each phase
    is shown for each vesicle class, along with the number of
    instances observed at either concentrations of oleic acid in the
    suspension (0.8 or 0.3~mM, pH~8.8).}
  \label{fig:classification}
\end{figure}

The changes in the vesicle morphology served as a basis for the
classification of the observations into groups.  Immediately after the
transfer into the oleic acid/oleate suspension, all vesicle transfer
recordings exhibited a period in which the initial spherical shape of
the selected vesicle did not change morphologically.  This phase was
followed by a phase of vesicle membrane evagination, where the vesicle
spawned several ``daughter'' vesicles.  Vesicles of the first group
remained in this phase throughout the course of the observation.  In
the second group, the initial evagination was followed by an
invagination, resulting in several invaginated vesicles within the
``mother'' vesicle.  In the third group, such an invagination was
followed by another evagination, forming several interconnected
spherical vesicles or thick tubular structures.  Denoting successive
evaginations by E and invaginations by I, we classified the recorded
vesicle transfers into five groups, from E to EIEIE. The sixth group
comprises the transfers discarded due to various reasons (no visible
changes; an abrupt vesicle rupture; the vesicle flew to a region where
recording was no more possible).  13 out of 54 transfers into 0.8~mM
oleic acid/oleate suspension and 2 out of 9 transfers into 0.3~mM
oleic acid/oleate suspension were discarded due to the problems
described.  Fig.~\ref{fig:classification} shows the classification of
the vesicle recordings into groups and indicates the frequency of
observations, together with examples of evaginated and invaginated
shapes.  No statistically convincing relationship has been found
between the class of the vesicle transfer and any measurable
parameter.  Possible influences include the volume and the
  area-to-volume ratio of the GUV, local variations in the
temperature, pH, and the convective flow in the chamber, as
well as the effects of micropipette manipulation during the transfer.
The recording was terminated when either the morphological changes
apparently stopped, or the changes were too slow to justify continuing
recording.

During the course of the experiment, no significant change in the
contrast between the GUV interior and exterior was observed,
\emph{i.e.}, throughout the experiment, the vesicle interior appeared
equally darker than the surrounding medium.  This indicates that the
amount of sucrose inside the vesicle which exchanged with the glucose
in the environment was below the detection level of our setup.  Since
the osmotic conditions remained unchanged, we can assume that the
volume of the GUV remained constant as well.

\begin{figure}
  \centering\includegraphics[width=0.45\textwidth]{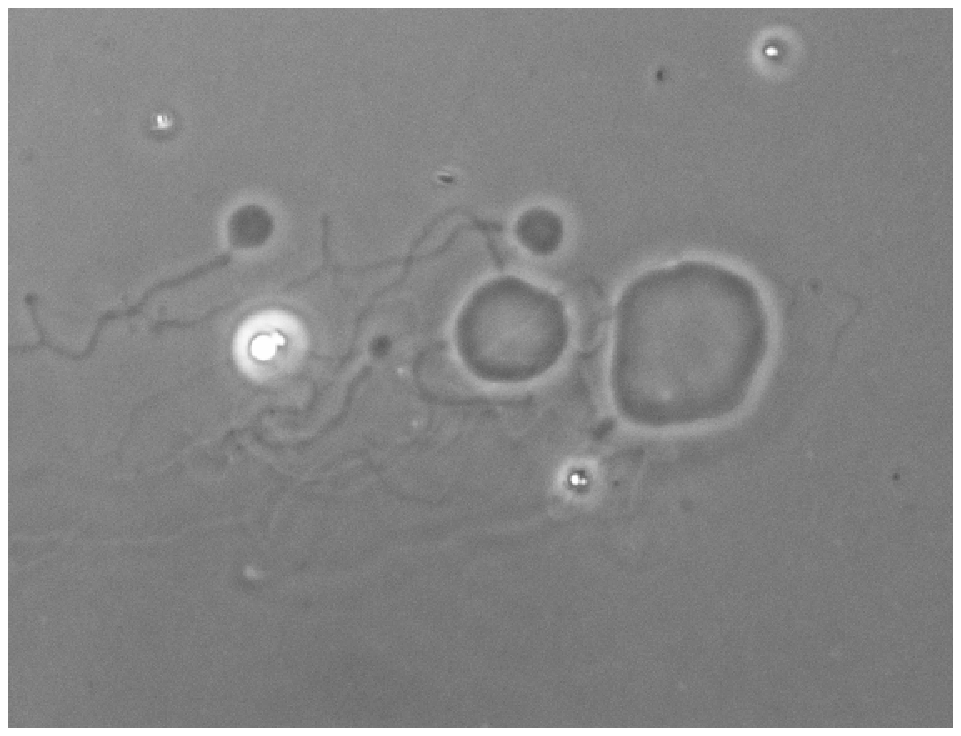}\\
  \vspace{\baselineskip}
  \centering\includegraphics[width=0.45\textwidth]{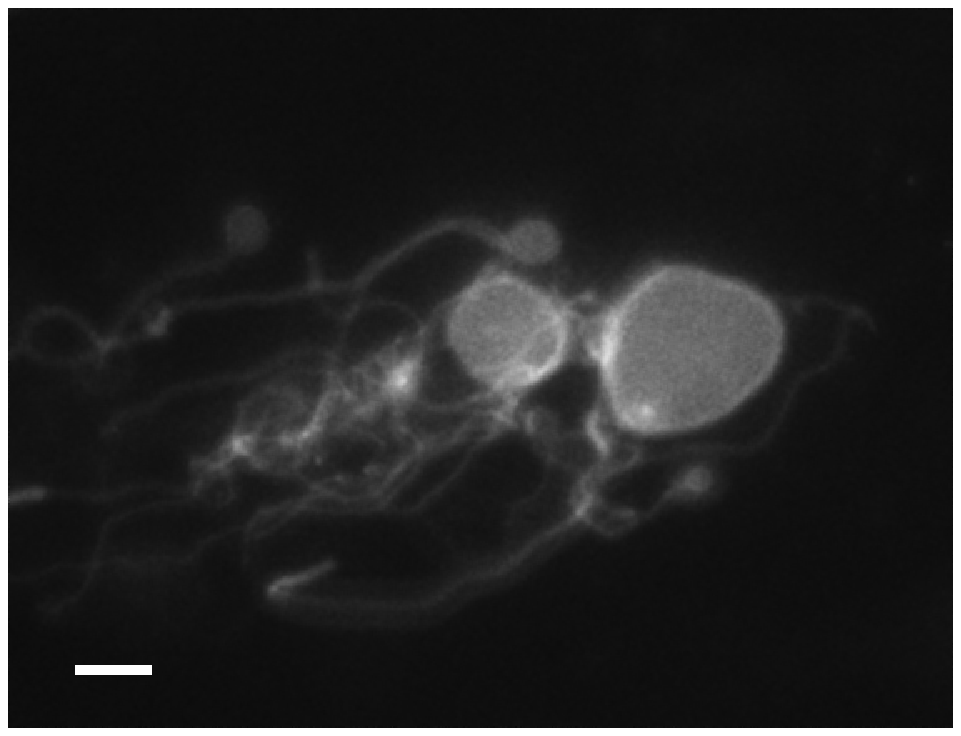}
  \caption{Top: a POPC GUV with 2\% NBD-labeled lipid shown
    approximately 3~min after the transfer into a 0.8~mM
    suspension of oleic acid/oleate in 0.2~M glucose (pH~8.8),
    shown with the phase contrast technique. Bottom: a fluorescent
    image of the same vesicle. The bar in the bottom frame represents
    10~$\mu$m.}
  \label{fig:fluoro}
\end{figure}

In several cases, the evaginations, which appeared upon the transfer
of a POPC GUV to either a 0.3~mM or a 0.8~mM suspension of oleic
acid/oleate, were in the form of thin tubular protrusions or tethers.
In order to test the presence of phospholipid in the thin tubular
protrusions, we conducted an experiment where POPC was marked with 2\%
NBD-labeled lipid.  We noticed fluorescence both in the tubular
protrusions and in the vesicle bodies (Fig.~\ref{fig:fluoro}),
and on the basis of the evidence that the
  calculated diffusion rates for NBD-labelled phospholipids
  \cite{Loura:2008} are comparable with the results for POPC
  \cite{Gaede:2003} (both $D \sim 10^{-11}\,\textrm{m}^2/\textrm{s}$),
  it is reasonable to expect that apart from NBD-PC, some unlabelled
  POPC may also be present in the tubular protrusions.

\subsection{Control experiments}
\label{subsec:control}

\begin{figure}
  \centering\includegraphics[width=0.48\textwidth]{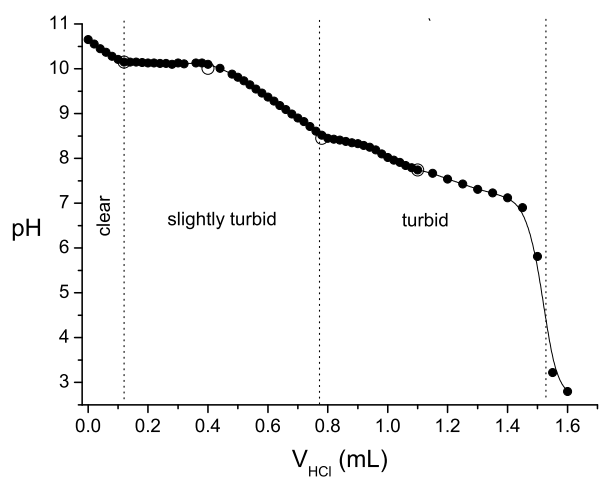}
  \caption{Equilibrium titration curve of 9.9~mM oleic acid/oleate
    (15~mL) determined with 0.1~M HCl at 25${}^\circ$C.}
  \label{fig:titration}
\end{figure}

The pH region where oleic acid/oleate at a total concentration of
9.9~mM forms bilayer structures was determined by titration.  The
titration was started with a 9.9~mM suspension of oleic acid
prepared in a 9.9~mM aqueous solution of NaOH, thus deprotonating
the COOH groups of oleic acid.  100~mM HCl was then added stepwise
and the pH was measured.  Fig.~\ref{fig:titration} shows the resulting
titration curve, which is in agreement with previously published
results obtained under similar experimental conditions
\cite{Cistola:1988,Fukuda:2001,Rogerson:2006} and indicates that
bilayer structures are present in the pH region from 8.4 to 10.2,
while above pH~10.2, only micelles exist, as indicated by the optical
transparency of the corresponding solutions.

\begin{figure}
  \begin{center}
    \begin{tabular}{cc}
      (a) & \includegraphics[width=0.45\textwidth]{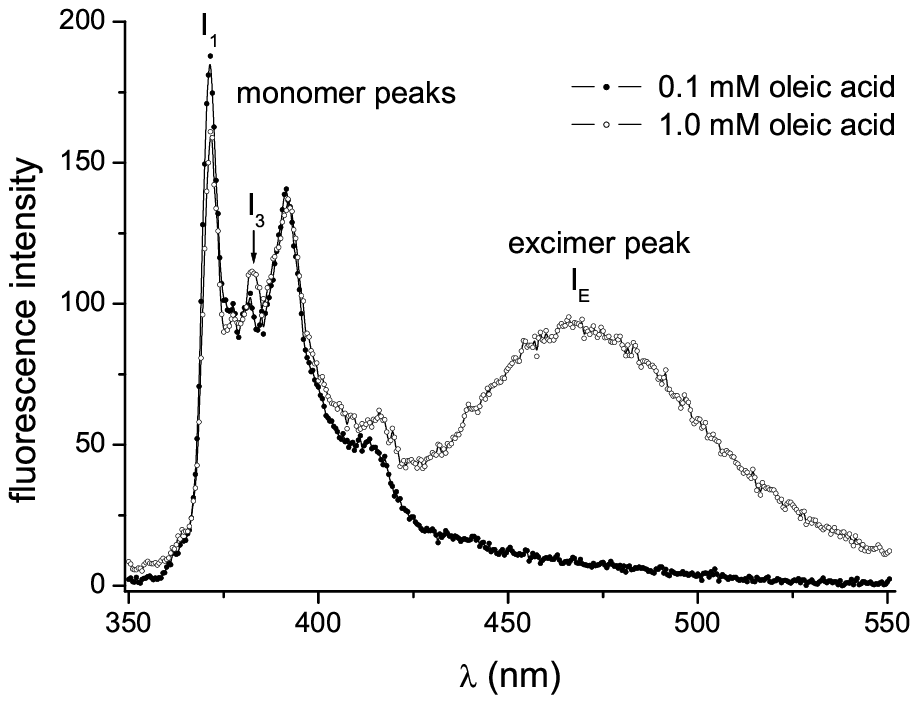}	\\
      (b) & \includegraphics[width=0.45\textwidth]{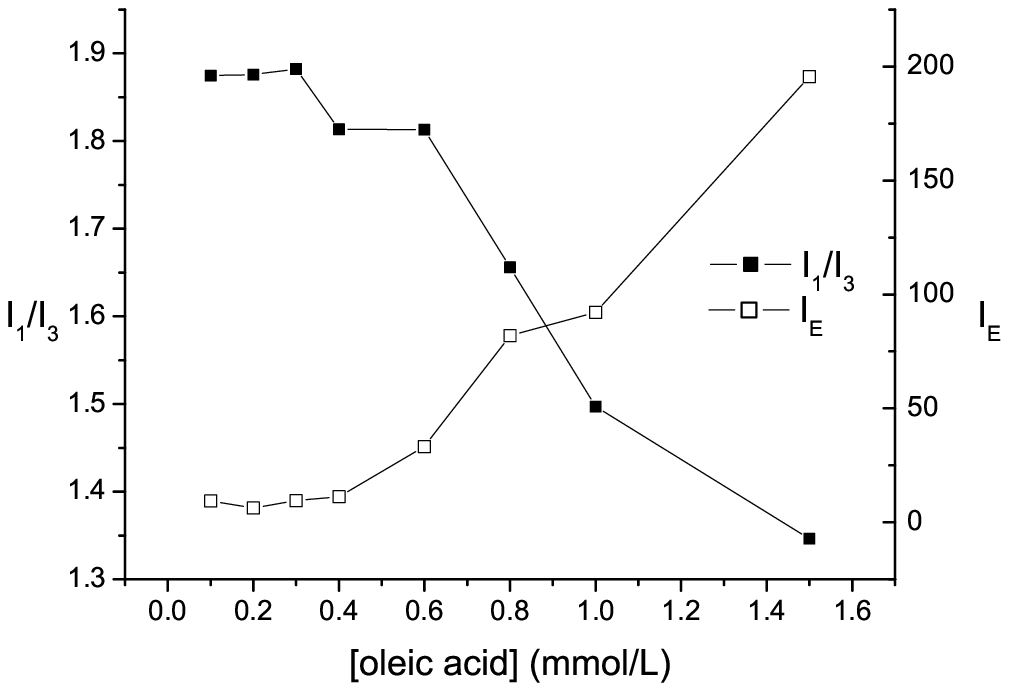}\\
    \end{tabular}
  \end{center}
  \caption{(a) Pyrene emission fluorescence spectra of 0.1~mM
    oleic acid ($\circ$) and 1.0~mM oleic acid ($\bullet$)
    buffered to pH~8.7.  Below the cvc, only monomer peaks are
    present, while above the cvc, a broad excimer peak appears.  (b)
    The intensity ratio of the monomer emission peaks $I_1/I_3$ and
    the intensity of the pyrene excimer peak ($I_E$; determined at
    473~nm) as a function of oleic acid concentration, buffered to
    pH~8.7.}
  \label{fig:spectrometry}
\end{figure}

The presence of oleic acid/oleate aggregates was determined using
pyrene fluorescence emission spectrometry.  It is known
\cite{Vanderkooi:1974,Almgren:1979} that pyrene is preferentially
solubilized in the hydrophobic environment of surfactant aggregates,
\emph{i.e.}, in the bilayer or micelle cores.
Fig.~\ref{fig:spectrometry}a displays the fluorescence emission
spectrum, showing that below the cvc of oleic acid/oleate
(0.4--0.7~mM, \cite{Walde:1994b}), only pyrene monomer peaks were
present, while a broad excimer peak appeared above the cvc, which is
consistent with published results \cite{Chen:2004b}.  The ratio of the
two monomer peaks, $I_1/I_3$, indicates whether pyrene molecules are
enclosed in a polar or a non-polar environment: high values (1.7 or
above) indicate that pyrene is in a highly polar environment
(\emph{e.g.}, water), while low values (1.3 or below) indicate a
non-polar environment \cite{Arrigler:2005}.
Fig.~\ref{fig:spectrometry}b shows the $I_1/I_3$ ratio and the height
of the pyrene excimer peak $I_E$, determined at 473~nm, as a function
of the concentration of oleic acid/oleate in suspension at pH~8.7.  A
simultaneous drop of $I_1/I_3$ and an increase of $I_E$ around
0.6~mM oleic acid in Fig.~\ref{fig:spectrometry}b indicates that a
non-polar environment started to be formed at this concentration.
This value is consistent with some of the estimates for the cvc in
this pH range \cite{Walde:1994b}.

In order to determine whether the presence of oleic acid/oleate
molecules is necessary for the observed shape changes, POPC GUVs were
transferred into a solution of glucose, buffered to pH~8.8 with
Trizma.  In 5 recorded vesicle transfers, no shape changes were
observed. 
Furthermore, if POPC GUVs were transferred into a suspension of POPC
LUVs in an iso-osmolar glucose solution, no shape changes occurred (8
recorded vesicle transfers).  This is in agreement with the
  previous findings \cite{Cheng:2003} and clearly shows that the shape
changes are related to the presence of oleic acid/oleate.

In an attempt to explain the consecutive transitions between an
evaginated and an invaginated shape, we wanted to exclude the
possibility that the evagination-to-invagination transition is a
simple effect of losing lipids from the outer membrane layer due to
adhesion of the membrane to the glass surface (the vesicles were in
contact with the bottom of the observation chamber, because they were
filled with a solution with a density greater than the density of the
environment).  In the experiment carried out, the POPC vesicle
membrane was marked with 1\% NBD-labeled lipid, and evidence for lost
patches of membrane during the vesicle movement over the surface due
to the convective flow was sought utilizing total-internal-reflection
fluorescence (TIRF) microscopy.  No evidence of lost patches of lipid
was found.

\subsection{Time course of a typical vesicle transfer}
\label{subsec:time-course}

Several characteristic events and selected stages of vesicle
development will be illustrated in the case of a single vesicle
transfer (Fig.~\ref{fig:Fig2} and a movie; see Supplementary data).  A
17 minutes long recording of a transfer of a vesicle from the group
EIE was selected as an example.

\begin{figure*}
  \begin{center}
    \begin{tabular}{cc}
      (a) & \includegraphics[scale=0.65]{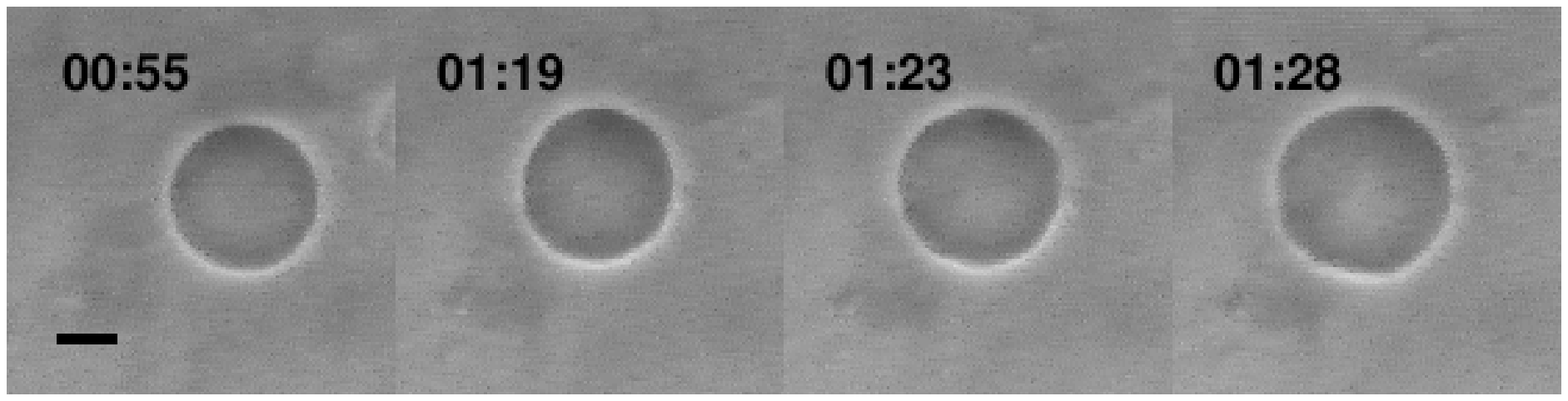} \\
      (b) & \includegraphics[scale=0.65]{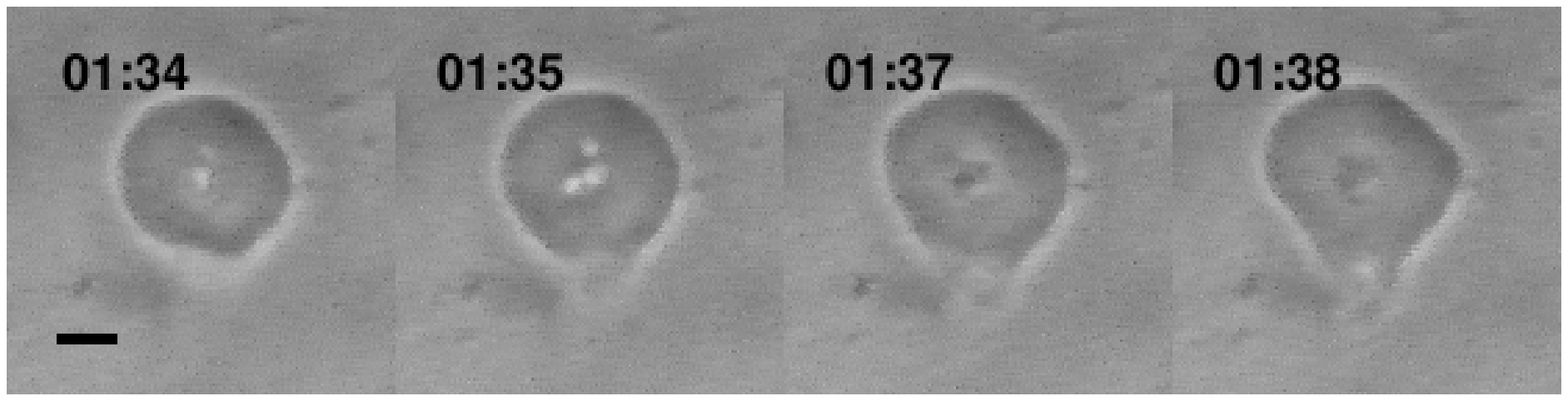} \\
      (c) & \includegraphics[scale=0.65]{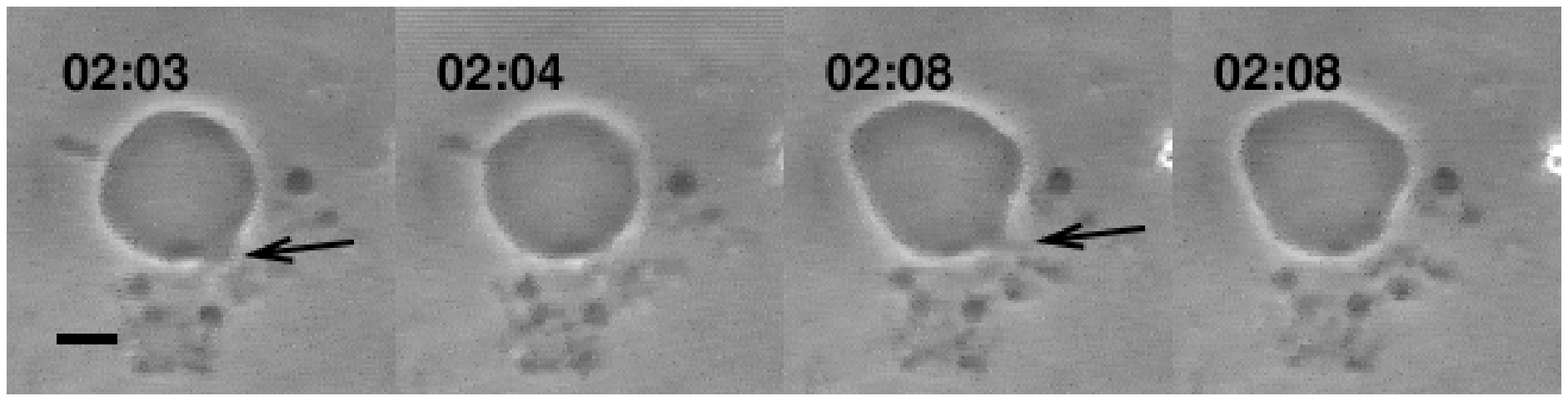} \\
      (d) & \includegraphics[scale=0.65]{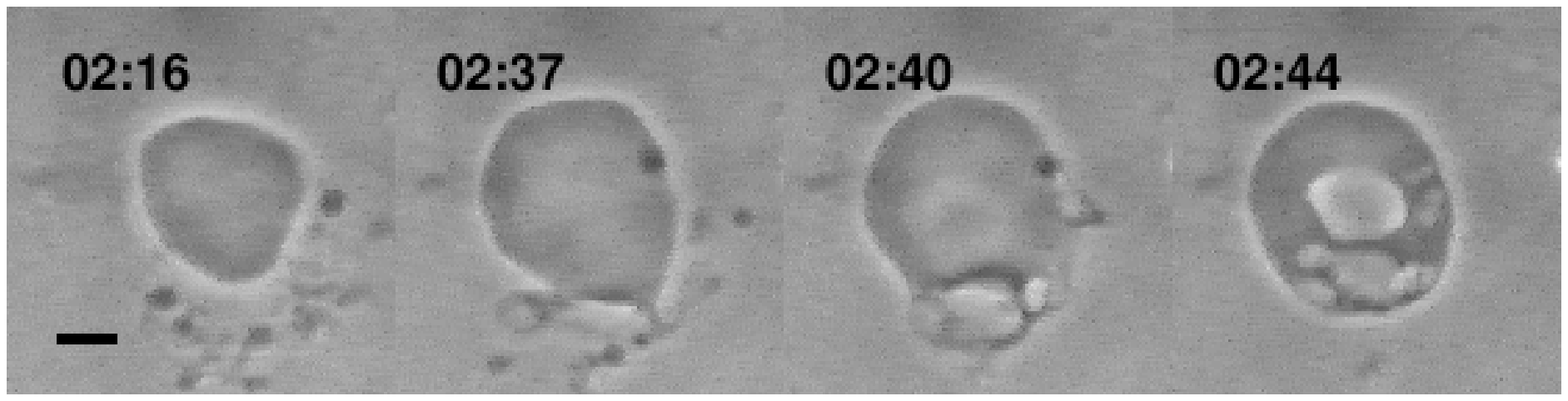} \\
      (e) & \includegraphics[scale=0.65]{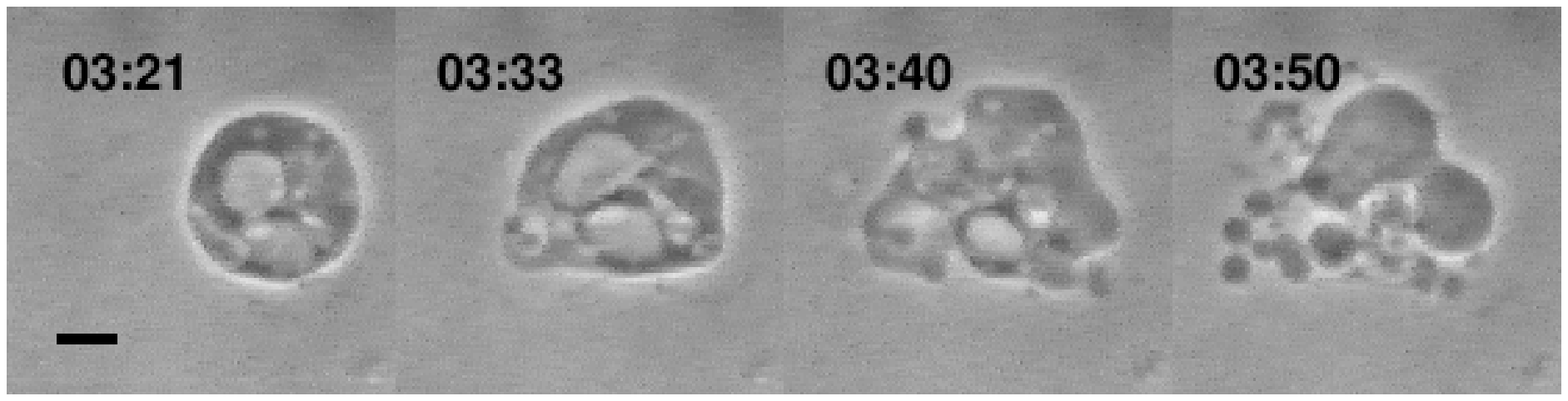} \\
    \end{tabular}
  \end{center}
  \caption{Characteristic events in the course of shape transitions of
    a POPC GUV upon the transfer into 0.8~mM suspension of oleic
    acid in 0.2~M glucose. (a) Initial increase of the vesicle
    cross-section. (b) An abrupt shape transition. (c) Vesicle budding
    (buds forming indicated by arrows). Two buds formed from a series
    of four consecutive buds are shown. (d) An invagination
    process. (e) An evagination process. The bar in the leftmost frame
    of each strip represents 10~$\mu$m. Time marks denote the time
    since the transfer.}
  \label{fig:Fig2}
\end{figure*}

\begin{figure}
  \centering\includegraphics[width=0.45\textwidth]{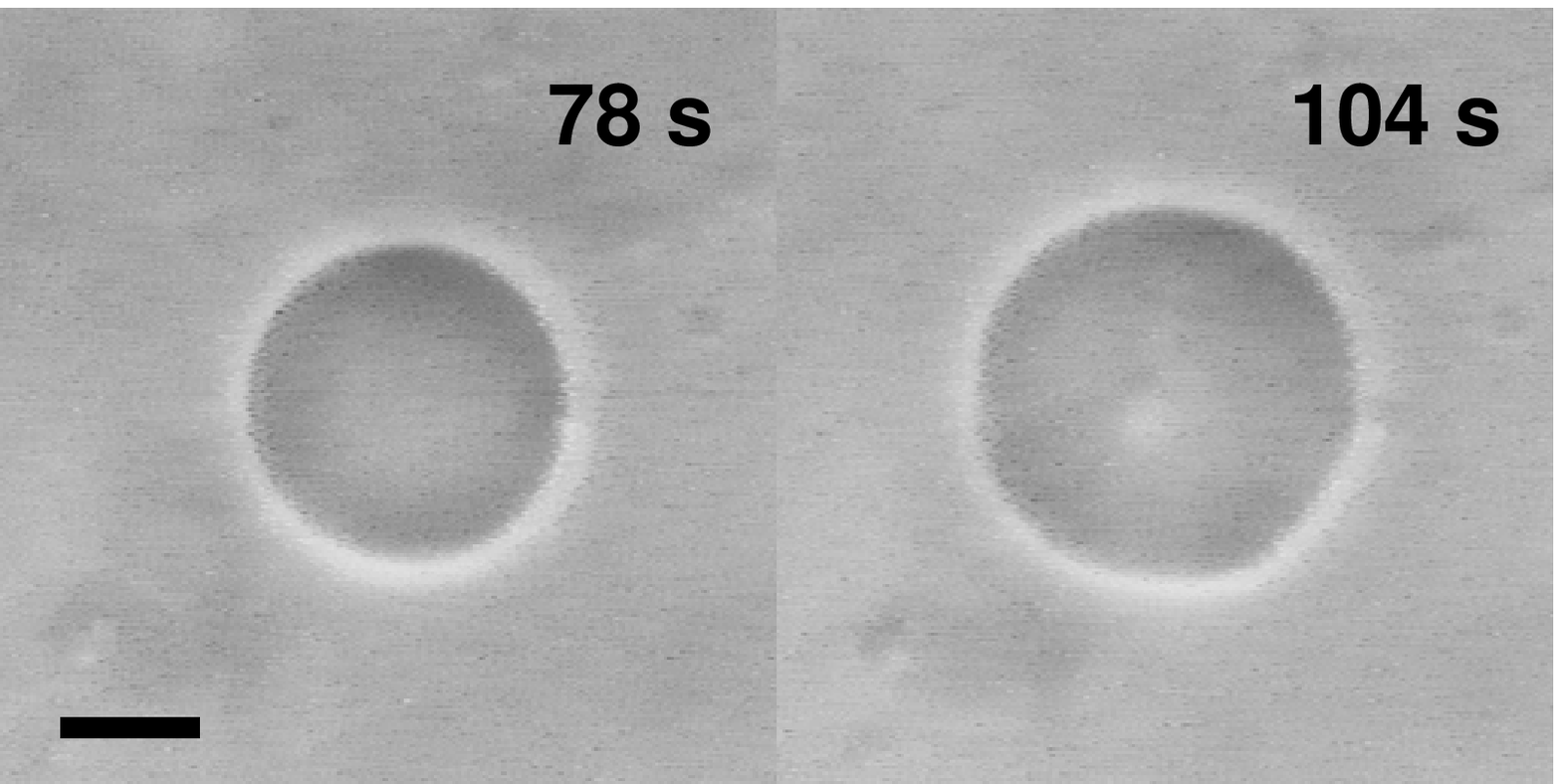}\\
  \vspace{\baselineskip}
  \centering\includegraphics[width=0.48\textwidth]{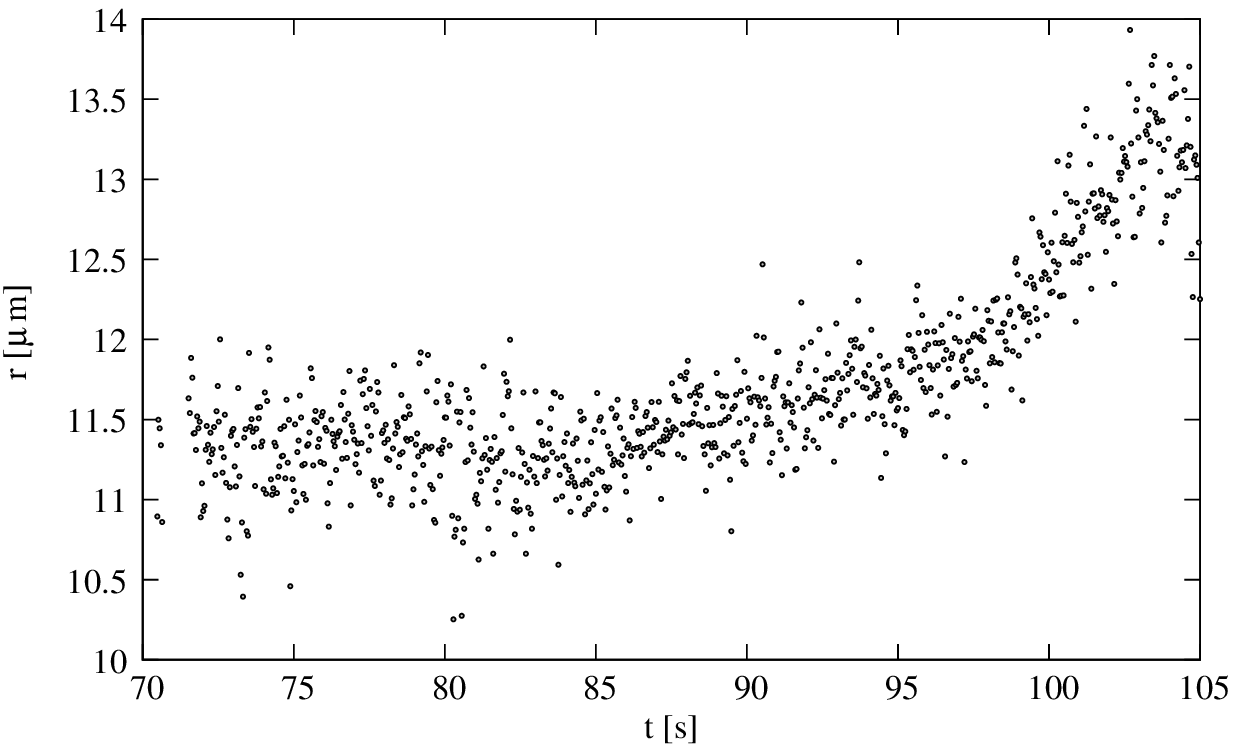}
  \caption{Top: a POPC GUV shown 78~s after the transfer into a
    0.8~mM suspension of oleic acid in 0.2~M glucose (left),
    and 104~s after the transfer (right). The bar in the left frame
    represents 10~$\mu$m. Bottom: vesicle radius as a function of the
    time elapsed after the transfer.}
  \label{fig:cross-section-increase}
\end{figure}

The increase of vesicle cross-section followed a characteristic
pattern (Fig.~\ref{fig:Fig2}a and movie, 01:00 to 01:28 (Supplementary data);
Fig.~\ref{fig:cross-section-increase}).  First, the vesicle sunk to
the bottom of the observation chamber within approximately 10~s.  The
sinking of the vesicle was verified by the necessary manipulation of
the microscope $z$-axis to keep the vesicle in focus, as well as by
the noticed decrease or even cessation of vesicle drift.  During the
time of sinking, as well as throughout a certain period after the
vesicle sunk to the bottom, the vesicle cross-section remained
spherical and its radius remained constant.  This phase was followed
by a phase when the vesicle cross-section was increasing at an
approximately constant rate while maintaining axial symmetry with
respect to the normal to the chamber bottom.

The phase of gradual increase of the vesicle cross-section was
terminated by an abrupt symmetry-breaking transition
(Fig.~\ref{fig:Fig2}b and movie, 01:33 to 01:38, Supplementary data),
where the vesicle spawned several daughter vesicles on the timescale
of seconds.  After this initial outburst, new daughter vesicles were
spawned one by one at an approximately constant rate.
Fig.~\ref{fig:Fig2}c (movie 01:55 to 02:13, Supplementary data) shows
two events when consecutive buds were forming in
a short time span on the same membrane section.  New daughter vesicles
were formed on the surface of the ``mother'' vesicle, while the
existing daughter vesicles were not affected.  It is worth noting that
the buds were approximately of the same size.  As it became evident in
the further development, all daughter vesicles remained connected to
the mother vesicle with narrow necks or tethers.

\begin{figure}
  \centering\includegraphics[width=0.45\textwidth]{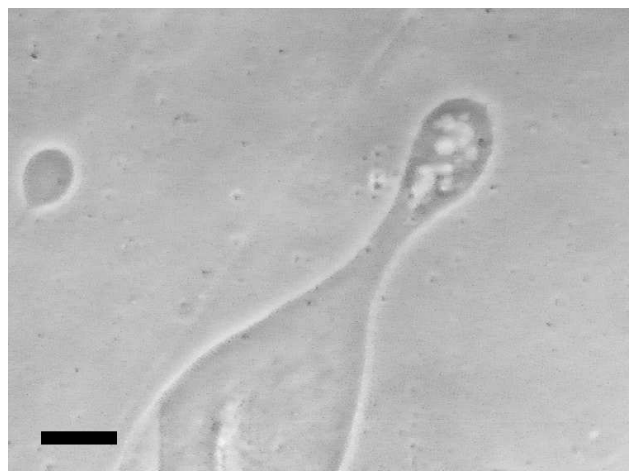}
  \caption{Invaginations start at the tip of protrusion, when the
    protrusion is retracting into the vesicle body. Image recorded
    2~min 7~s upon the transfer of a POPC GUV into a 0.8~mM
    suspension of oleic acid in a 0.2~M sucrose/glucose solution
    buffered to pH~8.8. The bar represents 20~$\mu$m.}
  \label{fig:invag-tip}
\end{figure}

After persisting in an evaginated shape for approximately 50~s, the
membrane abruptly started to invaginate (Fig.~\ref{fig:Fig2}d and
movie, 02:16 to 02:44, Supplementary data).  The complete
transition lasted approximately 20~s.  Invaginations often started at
the tip of the protrusions, which were then retracted into the vesicle
body (Fig.~\ref{fig:invag-tip}).

This invaginated phase ended after approximately 40~s, when the
membrane started to evaginate again (Fig.~\ref{fig:Fig2}e and movie,
03:20 to 03:50, Supplementary data).  The transition was completed
in approximately 25~s.  Afterwards, the vesicle remained in an
evaginated shape without qualitative changes of shape for the next 13
minutes (movie, 3:50--16:40, Supplementary data), after
which the recording was terminated.

\section{Discussion}

Although comparable earlier investigations performed on POPC LUVs
offer ample evidence that the oleic acid molecules are taken up by the
phospholipid vesicles, leading to their growth and eventually --
depending on the experimental conditions -- to vesicle fission
\cite{Lonchin:1999,Berclaz:2001a,Berclaz:2001b,Stano:2006}, no direct
real-time visualization of the process has been reported so far.  In
this work, instead of POPC LUVs, we used POPC GUVs since they allow a
direct observation by light microscopy and made some novel and
unexpected observations during the course of our investigations.  It
is the purpose of this discussion to offer a possible qualitative
explanation of the mechanisms underlying the observed phenomena: the
increase of the membrane area, the transient increase of the
difference of the areas of the outer and the inner membrane leaflet,
and the oscillations of the latter.  The section concludes by
indicating open issues, which require further work.

\subsection{Increase of membrane area}

\paragraph{Membrane area significantly increases}
On two occasions in the course of the experiment, an increase of the
membrane area can be demonstrated.  Shortly after the transfer, when
the POPC vesicle geometry is still simple, it is possible to estimate
its area with an accuracy of a few percent.  Towards the end of the
experiment, when the dynamics of the shape changes subsides, a more
crude estimate is possible.

The results plotted in Fig.~\ref{fig:cross-section-increase} show that
the diameter of the cross-section of an approximately spherical GUV
initially remained constant after it sank to the bottom of the
observation chamber with oleic acid/oleate suspension, and then
increased by about 15\%.  Approximating the transformation
with a change from a sphere to an oblate spheroid with a volume equal
to the volume of the sphere, one can estimate that a 15\% increase of
the diameter of the vesicle cross-section amounts to an approximately
3\% increase of the membrane area at this stage of the oleic
acid/oleate intercalation, which corresponds to oleic acid/oleate
representing $\sim$4--5 mol~\% in the membrane composition.  Flaccid
vesicles filled with a solution which exceeds in density the external
solution are known to deform into approximately oblate spheroids also
due to gravity \cite{Dobereiner:1997}.  However, the observed extent
of deformation can not be explained by gravitation alone.

The complex vesicle geometry towards the end of the experiment only
allows for a very rough quantification.  We estimate that the final
membrane area can be as large as five times its initial value, which
corresponds to oleic acid/oleate representing up to $\sim$85~mol~\% of
the membrane composition.  Vesicle shapes seem to reach an equilibrium
in the conditions of excess oleic acid/oleate: the amount of oleic
acid/oleate available in the chamber is $\sim 10^{17}$ molecules, and
thus vastly exceeds the amount of phospholipid: a 10~$\mu$m GUV
contains $\sim 4\times 10^9$ molecules, and even if 10- or 100-times
as much lipid is unintentionally introduced into the chamber with a
micropipette transfer, the ratio is still in the order of $\sim
1:10^6$.\footnote{It is assumed in the estimate that the average
  headgroup area of a phospholipid molecule is $a_0 \approx
  60$~\AA${}^2$ \cite[][Ch.~3]{CevcMarsh:PhospholipidBilayers} and
  that the vesicles are unilamellar; therefore a GUV with a radius $r$
  contains $N_\mathrm{PC} \approx 2\times 4\pi r^2/a_0$ molecules.}  A
significant increase of membrane area is in agreement with previous
experiments on POPC LUVs
\cite{Lonchin:1999,Berclaz:2001a,Berclaz:2001b}.

\paragraph{Membrane area growth can be related to system kinetics}
In a simple kinetic scheme describing the kinetics of the interaction
of fatty acids with phospholipid membranes \cite{Thomas:2002}, the
system is represented by the populations of oleic acid/oleate
molecules in the following four compartments: the vesicle exterior,
the outer membrane leaflet, the inner membrane leaflet, and the
vesicle interior.  Two of these, the populations of oleic acid in both
membrane leaflets are directly related to the equilibrium areas of the
outer ($A_{20}$) and the inner ($A_{10}$) membrane
leaflet.\footnote{We have used the index ``0'' to denote the
  equilibrium values of the area of a membrane leaflet, as opposed to
  their instantaneous values $A_2$ and $A_1$. The latter may differ
  from their equilibrium values: \emph{e.g.,} intercalation of
  molecules into the outer membrane leaflet, as in the experiment
  described, induces lateral tension in both membrane layers, the
  outer one being compressed and the inner one stretched.}  Both
$A_{20}$ and $A_{10}$ vary with time due to three processes: (i)
association of fatty acid molecules with the phospholipid membrane,
where the rate of the membrane area increase is proportional to the
concentration of oleic acid/oleate in the membrane vicinity and the
area available for intercalation; (ii) dissociation of fatty acid
molecules from the phospholipid membrane, its rate being proportional
to the oleic acid population in the membrane leaflet, and (iii) fatty
acid translocation (flip-flop) in the membrane, its net flow being
proportional to the difference of oleic acid populations in the two
leaflets.  Fatty acid association and dissociation occur both on the
external and the internal membrane-water boundary.  The association of
oleic acid monomers to a phosphatidylcholine membrane is considered to
be a very fast process \cite{Kleinfeld:1997,Cupp:2004}, while the
question whether the oleic acid translocation is faster
\cite{Kamp:1995,Simard:2008b} or slower
\cite{Kleinfeld:1997,Kampf:2006} than the dissociation appears to be
an unresolved question.

The demonstrated increase of the membrane area requires that the areas
of both membrane leaflets must increase, implying that by some means,
translocation of oleic acid must have occurred.  Furthermore, the
transitions cease once the oleic acid/oleate in the suspension is in
equilibrium with the oleic acid/oleate in the membrane on both the
internal and the external side of the membrane.  This implies that the
concentration of oleic acid/oleate in the vesicle lumen is equal to
the external concentration, and that both the translocation and
dissociation must have occurred within the observed timescale.

\subsection{Transient increase of the difference between the areas of
  the outer and the inner membrane leaflet}

\paragraph{Shape changes commence with an evagination}
In all 48 observed cases of POPC GUV transfer into a suspension of
oleic acid/oleate, its concentration being either above or below the
cvc, shape transformations always commenced with a transition into an
evaginated shape.  In 16 out of 41 cases where the concentration of
oleic acid/oleate in the suspension was above cvc and in all 7
observed cases where the concentration of oleic acid/oleate in the
suspension was below the cvc, vesicle persisted in this shape until
the end of the experiment.

\paragraph{The relation with the bilayer couple model}
Vesicle shape can be related to the difference between the areas of
the outer and the inner membrane leaflet $\Delta A_0 = A_{20}-A_{10}$
\cite{Svetina:1992}.  Given the values of $A_0 = 2
A_{10}A_{20}/(A_{10}+A_{20})$ and $\Delta A_0$, a vesicle adapts its
shape (along with $A$ and $\Delta A$, which correspond to a given
shape) in order to minimize its total energy, comprising of membrane
stretching and membrane bending energy \cite{Helfrich:1973}, as well
as the entropic free energy contribution due to thermal fluctuations
\cite{Seifert:1997}.  Shapes with $\Delta A_0$ larger than the value of
$\Delta A_0$ for a sphere are evaginated, and those smaller are
invaginated.  At values of $\Delta A_0$ which significantly exceed the
value of $\Delta A_0$ for a sphere, it is expected that vesicle shapes
look like the so-called limiting shapes, \emph{i.e.}, shapes with a
maximal volume at given $A$ and $\Delta A$, which were shown to
consist of connected spheres of two allowed sizes
\cite{Svetina:1989,Kaes:1991,Farge:1992}.  The observed
(quasi-)equilibrium shapes were one or a few bigger spheres connected
to numerous smaller spheres, which approximately agrees with the
bilayer couple model.  This state may superficially resemble vesicle
growth and fission observed in the experiments with LUVs (\emph{e.g.},
\cite{Blochliger:1998,Lonchin:1999,Berclaz:2001b,Stano:2006}),
however, we need to emphasize that in our case the newly formed
daughter vesicles remained tethered to the mother vesicle,
\emph{i.e.}, no real separation of membrane into two entities has been
observed.

\paragraph{The effect of oleic acid intercalation on spontaneous curvature}
Apart from affecting $A_0$ and $\Delta A_0$, the intercalation of
oleic acid molecules into POPC membrane changes its composition. This
could also affect its spontaneous curvature \cite{Helfrich:1973},
since the spontaneous curvature depends on the shapes of the membrane
constituents.  POPC molecules are approximately cylinder-shaped, which
makes their intrinsic curvature close to zero, favoring their
aggregation into flat bilayer structures \cite{Israelachvili:I&SF}.
Protonated oleic acid molecules have a slightly negative intrinsic
curvature \cite{Bergstrand:2001}, while deprotonated oleic acid
molecules, \emph{i.e.}, oleate ions, exert mutual electrostatic
repulsion between their headgroups, which favors their aggregation
into micelles or other structures with highly positive curvature.  At
$\textrm{pH} \sim \textrm{pK}_a$, where approximately half of the
oleic acid molecules are protonated, it is known that the aggregates
formed by oleic acid/oleate molecules are bilayers.  If one assumes a
similar behavior in mixed phospholipid/oleic acid/oleate bilayers, the
average spontaneous curvature of a membrane composed of phospholipid,
and equal amounts of oleic acid and oleate is expected to be close to
zero.  We can therefore conclude that in the conditions the experiment
was conducted, \emph{i.e.,} at $\textrm{pH}\sim\textrm{pK}_a$, oleic
acid/oleate forms bilayer structures with zero intrinsic curvature,
and the effect of oleic acid intercalation into the membrane onto the
spontaneous curvature is minor in comparision to its effect on $\Delta
A_0$.

\paragraph{Consecutive vesicle budding}
On several occasions, usually later in the course of morphological
changes caused by the incorporation of oleic acid into the membrane,
we noticed several consecutive buds being formed on the same membrane
section in a short time span (Fig.~\ref{fig:Fig2}c; movie 01:55 to
02:13, see Supplementary data).  Such a process, which lasted for a
certain period of time, means that during this period, $\Delta A_0$
and $A_0$ were increasing in the appropriate ratio.  Such processes
have not been observed in the experiments with lysolipids
\cite{Needham:1995,Zhelev:1996}, because in the latter case, no
significant change of $A_0$ has been noticed.  It is worth noting that
the buds were usually found to be of approximately the same size,
which is a confirmation that the vesicle shape belongs to limiting
shapes.  The budding is also an important mechanism in the processes
of membrane trafficking, \emph{e.g.}, in the endoplasmatic reticulum
and endocytosis.  The studies of budding in cells indicate that this
process can only occur with the support of a complex protein network
\cite{Watson:2005}.  The observed consecutive formation of buds during
a certain period of time demonstrates that budding can also occur as a
simple physical reaction of membrane to a simultaneous increase in
both the total membrane area and the difference in the outer and the
inner leaflet area.

\subsection{Oscillations of $\Delta A$}

\paragraph{Non-specific convective flow}
In 25 out of 41 cases of vesicle transfer into a suspension where the
concentration of oleic acid/oleate is above cvc, the initial membrane
evagination has been followed by a membrane evagination.  Such shape
changes correspond to a decrease of $\Delta A$ from a value larger to
a value smaller than the value of $\Delta A$ for a sphere.  This
phenomenon can not be explained with a simple kinetic model mentioned
above.  In addition to oleic acid association, dissociation and
diffusive translocation, we propose another non-specific mechanism for
translocation, which employs membrane defects or pores that occur
stochastically \cite{Raphael:1996,Raphael:2001}.  In the initial stage
shortly after the transfer of GUV, oleic acid intercalation into the
outer membrane leaflet leads to the overpopulation of the outer
membrane leaflet with respect to the inner one, and consequently to an
imbalance in lateral tensions of the two membrane leaflets.  The two
membrane leaflets tend to relax, and when a membrane defect occurs,
the difference in lateral tensions causes a convective flow of
membrane components from the outer to the inner leaflet.  It is to be
noted that all species of molecules -- phospholipid, oleic acid and
oleate ions -- can traverse the membrane via the convective flow,
their ratios in the flow being proportional to their ratios in the
outer membrane leaflet.  Via this mechanism, the population of
phospholipid in the inner membrane leaflet exceeds its population in
the outer membrane leaflet.  At this point, phospholipids alone would
favour invaginated shape; however, due to the influx of oleic
acid/oleate into the outer membrane leaflet, the overall shape is
still evaginated.  As the diffusive flip-flop rate for oleic acid is
much higher than the corresponding flip-flop rate for phospholipids
\cite{John:2002,Liu:2005}, at some point both the concentration of
oleic acid/oleate in the vesicle lumen equilibrates with the external
concentration, and the population of oleic acid/oleate in the inner
membrane leaflet equilibrates with the one in the outer membrane
leaflet.  A low flip-flop rate for phospholipids together with
virtually no net convective flow, since there is no imbalance in
lateral tensions of the two membrane layers either, means that this
quasi-equilibrium state is a kinetic trap for the phospholipid.  With
oleic acid in balance, the vesicle shape is determined by the
populations of phospholipid in the two membrane leaflets, which favour
an invaginated shape.  Note that the vesicle interior is osmotically
balanced with the exterior throughout the experiment, so no net flux
of water is expected to occur through these pores.  We can thus
conclude that the observed shape transition from evaginated to
invaginated shapes can be explained, if one allows for a non-specific
convective flow through stochastically occurring membrane defects
\cite{Raphael:2001}.

\subsection{Open issues}

\paragraph{A burst of satellite vesicles vs.\ a gradual increase in
  their number}
In the experiments carried out, we observed an abrupt transition,
where GUV persisted in an approximately oblate shape for a certain
period of time, this being followed by sprouting a number of satellite
vesicles in a short period of time.  This is seemingly in
contradiction with the bilayer-couple theory, which predicts a gradual
one-by-one increase as the $\Delta A_0$ increases.  The explanation
for this absence of intermediate steps possibly involves the entropic
contribution due to thermal fluctuations \cite{Seifert:1997}, which
lowers the free energy of a fluctuating oblate spheroid below the free
energy of a spherical vesicle with spherical satellites.

\paragraph{The satellite spheres were not all exactly the same size}
The observed state with a few larger vesicles and several smaller
satellite vesicles of approximately the same size is consistent with
the notion of the bilayer-couple model.  However, the satellite
vesicles observed in the experiment were not exactly of the same size,
and there are several possible reasons for that: (a) the model is
valid for an equilibrium system, while the observed system might not
have reached the equilibrium yet, (b) the model is valid for a
homogenous membrane, while the mixed phospholipid/oleic acid membrane
is not necessarily laterally homogeneous, and (c) satellite vesicles
formed at different times are also formed at slightly different
conditions (\emph{e.g.,} the composition of membrane changes);
however, once formed, they are found in a kinetic trap, as the energy
needed to reopen and close the neck connecting two spheres far exceeds
the energy difference between different configurations.

\paragraph{Higher oscillations} We observed a comparably frequent (11
out of 41 cases) shape transitions from invaginated shape back into
evaginated shape, and in a few cases even higher oscillations
(Fig.~\ref{fig:classification}).  Although it remains to be seen
whether the convective-flow mechanism described above can reproduce
such behaviour with realistic values of parameters, we would not want
to exclude the possibility that another mechanism is in effect,
possibly connected to spatial inhomogeneity of oleic acid/oleate
concentration.

\paragraph{Direct interaction with oleic acid/oleate aggregates}
A qualitative difference between the GUV behaviour in the cases where
the concentration of oleic acid/oleate in the suspension was below
cvc, and the cases where its concentration was above cvc, \emph{i.e.,}
where oleic acid aggregates were present, opens a possibility that
another interaction pathway was present in the latter case.  While the
prevalent opinion seems to be that the oleic acid/oleate monomers
dissociate from the donor system before being incorporated in the
phospholipid membrane (\emph{e.g.,} \cite{Thomas:2002}), some authors
allow for a possibility that oleic acid/oleate aggregated
intermediates directly interact with a phospholipid membrane
\cite{Chen:2004b}.  The two proposed pathways differ in their effect
on the vesicle shape.  While the interaction with monomers directly
increases the area of the outer membrane leaflet and only indirectly,
through translocation, the area of the inner membrane leaflet, a
possible interaction with an aggregated intermediate would contribute
to both membrane leaflets at the same time.  However, since LUVs have
a large intrinsic $\Delta A_0$, both mechanisms are expected to result
in a shape evagination, and the experiment we conducted can not be
used to distinguish between the two.

\paragraph{Oleic acid flip-flop: fast or slow?}
There is an ongoing debate whether the oleic acid translocation is
fast or slow.  The results obtained with free fatty acids
\cite{Kamp:1995,Simard:2008b} yield much higher rates than those
obtained with BSA-complexed fatty acids
\cite{Kleinfeld:1997,Kampf:2006}.  It has been suggested
\cite{Cupp:2004} that exposure to fatty acid concentrations higher
than 5~$\mu$M somehow perturbs the bilayer structure, which may result
in a higher apparent translocation rate.  We would like to emphasize
that the two mechanisms for oleic acid/phospholipid translocation
proposed in this paper -- diffusive flip-flop and non-specific
convective flow -- work in a similar way.  At very low concentration
of oleic acid/oleate in the suspension, there is virtually no
imbalance in lateral tensions of the two membrane layers, and the
outward and the inward convective flows are balanced, thus the only
mechanism for oleic acid translocation is diffusive flip-flop, driven
by the imbalance in oleic acid concentration.  At higher
concentrations, however, convective flow, driven by the imbalance in
lateral tensions, opens another, possibly faster, pathway.  Although
this complex issue is probably far from being resolved, we hope this
notion may help by providing another viewpoint to this seemingly
conflicting issue.

\section{Conclusions}

By working with GUVs in the optical microscopy regime, this study
offers a new insight into the budding process by which new daughter
vesicles can be formed from phospholipid vesicles by virtue of the
incorporation of fatty acids into the phospholipid membrane.  These
observations complement previous findings obtained by other methods,
which showed an increase of the number of LUVs upon the transfer into
an oleic acid/oleate suspension \cite{Stano:2006}.  We offer a
qualitative explanation of the observed phenomena, however, further
work in the direction of the determination of kinetic and mechanical
properties of this system and building the corresponding theoretical
model will be needed for its confirmation.  The observed budding and
formation of satellite vesicles induced solely by a membrane growth
without the intervention of proteins may have an implication for the
understanding of the basic principles governing membrane trafficking.

\section*{Acknowledgments}

The authors would like to thank Dr.~Bojan Bo\v{z}i\v{c}, Dr.~Martin
Hanczyc, Dr.~Janja Majhenc, and Dr.~Andrea Parmeggiani for their
valuable discussions and suggestions.  Dr.~Pegi Ahlin Grabnar
performed the photon correlation spectroscopy of oleic acid/oleate
vesicles, and Miha \v{S}u\v{s}ter\v{s}i\v{c} helped with the
  vesicle classification.  This work has been supported by the
Slovenian Research Agency research grant P1-0055 and the COST Action
D27.

\section*{Supplementary data}

An annotated video recording from which the strips in
Fig.~\ref{fig:Fig2} were selected is available as an online
supplement. This information is available  via the
Internet at \url{http://biofiz.mf.uni-lj.si/~peterlin/supplement1.avi}.

\end{document}